\documentclass{article}		
\usepackage{latexsym}		
\begin{document}

\def\vector#1{\vec{\bf {#1}}}       
\def\baseE#1{{\bf E}_{#1}}
\def\basise#1{{\bf e}_{#1}}
\def\Tau{\tau}      
\def\eq#1{eq.\ ({#1})}
\def\BAR#1{\overline{#1}}
\def\Ed#1{{\bf E}_{#1}}
\def\Edd{{\bf e}_1 \wedge {\bf e}_2}
\def\Eda{{\bf E}_0}
\def\Edb{{\bf e}_1}
\def\Edc{{\bf e}_2}
\def\HALF{{1 \over 2}}

\begin{titlepage}

\title{\Large \bf Polydimensional Relativity, a Classical Generalization
of the Automorphism Invariance Principle
\thanks{FTP://www.clifford.org/clf-alg/preprints/1996/pezz9601.latex}}

\author{{\bf William M. Pezzaglia Jr.}
\\Department of Physics\\
Santa Clara University\\
Santa Clara, CA 95053\\U.S.A.\\
Email: wpezzaglia@scuacc.scu.edu}
\date{(Received: June 1994)}
\maketitle
\thispagestyle{empty}


\begin{abstract}
The {\it automorphism invariant} theory of Crawford\cite{Crawford}
has shown great promise, however its application is limited by the
paradigm to the domain of spin space.  Our conjecture is that there
is a broader principle at work which applies even to classical physics.
Specifically, the laws of physics should be invariant under 
polydimensional transformations which reshuffle the geometry (e.g.
exchanges vectors for trivectors) but preserves the algebra.
To complete the symmetry, it follows that the laws of physics must
be themselves polydimensional, having scalar, vector, bivector etc.
parts in one multivector equation.  Clifford algebra is the natural
language in which to formulate this principle, as vectors/tensors were for
relativity.    This allows for a new treatment of the relativistic
spinning particle (the Papapetrou equations)
which is problematic in standard theory.  In curved
space the rank of the geometry will change under parallel transport,
yielding a new basis for Weyl's connection and a natural
coupling between linear and spinning motion.

Note: Summary of talk, to appear in:
{\it Proceedings of the 4th Conference on Clifford
Algebras and their Applications in Mathematical Physics, Lehrstuhl II
f\"ur Mathematik, Aachen, Germany, May 1996}, K. Habetha editor.
\date{}
\end{abstract}
\end{titlepage}
\pagestyle{myheadings}

\section*{I. Introduction}
There has
been relatively few new physical principles proposed which are based
upon the unique structure of {\it geometric algebra}.  A notable exception is
the form of {\it spin gauge theory} put forth by Crawford\cite{Crawford}.
His proposition is that quantum mechanics should be form invariant
under local changes in spinor space basis (equivalently the 
matrix representation of the Dirac algebra can be different at each
point in space).  The motivation is to have a unified theory in
which the gauge fields of curvature describe gravity
as well as all the other fundamental forces.

The action of this local metric-preserving {\it automorphism 
transformation} is to ``mix up''
the basis elements of the full Clifford algebra, such that the basis
vector generators $\gamma_\mu$
at one point could be a mixture of the bivector, trivector, etc.
at another point.  However, a reshuffling of this ``spin'' geometry
$\gamma_\mu$  (i.e. the ``soldering forms'' which connect the
spinor basis to the tangent basis of spacetime) 
will \underline{not} 
change the physical basis vectors $\basise{\mu}$ of real geometric
spacetime into something other than a vector.  The two algebras are
independent; any element of the ``spin'' Clifford algebra 
$\gamma_\mu$ will necessarily commute\cite{PezzMex} with the 
basis vectors $\basise{\alpha}$.  In order to get the curvature
of the spin space to ``create'' curvature in coordinate spacetime,
a constraint must be imposed by fiat.
For example, in {\it general relativity}, the condition that the 
covariant derivative of the metric tensor will vanish is equivalent to
stating that the universe has the geometric structure of a Riemann space.
In {\it spin gauge theory}, the different constraint choices imposed by various
authors (usually obscured in some reasonable sounding assumption)
is making some sort of classification of the type of spin space plus
geometry space in which unified phenomena exists.

The most unambiguous way to choose the connection between spinor space
and coordinate space is to simply have one unified geometric
language for classical fields and quantum mechanics.
Column spinors are replaced by geometric multispinors (aggregates of
scalar, vector, bivector, etc.) which are left ideals of the
algebra\cite{Pezz592}. 
Now the Dirac matrices $\gamma_\mu$ can be varying
linear combinations of \underline{only} the basis vectors
$\basise{\alpha}$ at each point in space, with necessarily vanishing
covariant derivative (whereas Crawford has it to be non-vanishing).
The general automorphism transformation must be disallowed because it
would reshuffle the \underline{full} spin algebra.  Except for electromagnetic
and gravitational fields, all of Crawford's interesting features are
necessarily suppressed.  In order to describe other interactions,
Chisholm and Farwell\cite{Chisholm} resort to introducing higher
dimensions to the Clifford algebra (at last count 7 extra dimensions
on top of the 4 of spacetime).

Our own approach has been to stay within the 4D algebra, but make use
of all 16 geometric degrees of freedom in the multivector wavefunction
to describe multiple generations of particles\cite{Pezz592}.
In order to accomodate all the known couplings, we were heuristically
led to consider a new form of {\it bilateral} (left and right sided)
multiplication on the wavefunction that can not be derived from a
gauge transformation.  The action of this operation is
equivalent to a linear transformation on the full Clifford algebra, and
hence can be cast into a form which resembles automorphism gauge theory.
The problem is that Crawford's principle\cite{Crawford} is limited by
the paradigm to spin space.  We take a big leap and propose that classical
physics obeys the automorphism principle.  This has broad consequences
to both special and general relativity, some examples of which are explored
in the following sections.

\section*{II. Extension of Special Relativity}
Einstein required the laws of physics to be invariant under
Lorentz transformations, which ``rotate'' between scalar time and
vector space.  We propose a generalization: {\it the laws should be
invariant under Automorphism transformations} which reshuffle vector
space with bivectors, trivectors, etc.

\subsection*{A. Review of Standard Theory}
According to Minkowski, the world is a four dimensional continuum, which
we often call {\it spacetime}.  Events $\Sigma$ are points in the manifold
with coordinates $(t,x,y,z)$, where the fourth dimension is ``time''.
One of the postulates that Einstein put forth is that
{\it the speed of light is the same for all observers};
equivalently the speed of light
``$c$'' is a physical limit which cannot be exceeded.  Geometrically this
forces the metric measure of time to be the opposite sign as the other
dimensions, such that distance $d\Sigma$ between two points is measured
as the root of,
$$c^2 d\tau^2 = c^2 dt^2 - (dx^2+dy^2+dz^2).  \eqno(1)$$
The affine parameter $\tau$ is commonly called the {\it proper time}.
The other postulate upon which relativity is based is that
{\it the laws of physics are invariant in all inertial (nonaccelerated)
frames}.  
Specifically this means that
physical formulations must be the same in reference frames which differ
only by constant velocity; equivalently formulas [such as \eq{1}]
must be invariant under the Lorentz group $SL(2,{\bf C})$.

The {\it principle of least action} states that
a particle will ``choose'' to take the
path of least distance (in spacetime).
Using the calculus of variations, one minimizes the action integral,
based upon \eq{1},
$${\cal A}=\int{{\cal L} d\tau}=\int{m_0c d\tau}=
\int{m_0 c \sqrt{u^\alpha u_\alpha}\ d\tau},
\eqno(2) $$
which is clearly invariant under the Lorentz group.
The integrand ${\cal L}$ is called the {\it Lagrangian}, which is
generally a function of the coordinates $x^\alpha$ and the velocities
$u^\alpha=\dot x^\alpha =dx^\alpha /d\tau$.
The {\it four-momentum} $p^\mu$,
$$p^\mu={\delta {\cal L} \over \delta u^\mu}=m_0 u^\mu, \eqno(3a)$$
is conserved in time.  In addition to having one more component,
it differs from the
non-relativistic three-momentum $\vector{P}=m\vector{v}$
by the  {\it Lorentz factor} $\gamma$,
$$\gamma = {dt \over d\tau} = 
{u^0 \over c}=\left({\displaystyle
 \sqrt{1-{v^2 \over c^2}}}\right)^{-1}, \eqno(3b)$$
which appears in kinematic formulas as the
``relativistic correction'' (e.g. mass increases by: $m=\gamma m_0$).

Non-relativistically, rotational motion is ``uncoupled'' from
the linear motion.  This is not the case in relativistic theory
where the Pauli-Lubanski spin polarization four-vector $s^\mu$ must
everywhere be perpendicular to the four-momentum:
$p_\mu s^\mu = 0$,
(known as the Dixon\cite{Dixon} transversality condition, other authors
use the slightly inequivalent Frenkel\cite{Frenkel}
condition: $\dot x_\mu s^\mu=0$).
Hence if the linear motion changes with time, so must the spin.
One can argue for reciprocal effects.  When a particle is boosted in
a direction perpendicular to its spin, the mass on one side is moving
faster than on the other, causing an asymmetric relativistic mass
distribution resulting in a sideways shift of the center-of-mass.
Under either linear or angular acceleration this causes a sideways
contribution to the momentum.  Hence the conserved momentum is no
longer parallel to the velocity,
$$p^\mu = m \dot x^\mu + \dot S^{\mu\alpha}\ \dot x_\alpha, \eqno(4a)$$
$$p^\mu\ \dot x_\mu = -m_0 c, \eqno(4b)$$
$$S^{\mu\alpha}={1 \over 2} \epsilon^{\mu\alpha\beta\delta}
\ p_\beta \  s_\delta, \eqno(4c)$$
$$\dot S_{\mu\nu} = \dot x_\mu p_\nu -\dot x_\nu p_\mu. \eqno(4d)$$
There is some disagreement over the proper form of these equations
(we have followed Barut\cite{Barut}).  Interestingly, the equations
of motion appear to admit self-substaining circular solutions with
no net momentum, for which there are various possible physical
interpretations.  This feature may be an artifact of the 
coordinates no longer being a true description of the center-of-mass
of a spinning particle.  Regardless,
the problem at hand is that it is difficult to find a generalization
of \eq{2} which will simultaneously give both the
equation of motion for the translation and the spin.  A recent review
of the various methods is given by Frydryszak\cite{Frydryszak}.

\subsection*{B. The Clifford Manifold}
We propose that space is a fully {\it polydimensional} continuum.
Each event $\Sigma$ is a generalized ``point'' in a Clifford 
manifold which has a coordinate
$ q^A $ associated with \underline{each} basis multivector
element $\baseE{A}$ of the geometry.  As an example, consider a disk
(hockey puck) constrained to move on a 2D (flat) Euclidean surface.
The set of basis elements $\{ \baseE{A} \}$ generated by two
anticommuting basis vectors is: $\{ \baseE{0}, \baseE{1},
\baseE{2}, \baseE{3} \} = \{ {\bf 1},\basise{1},\basise{2},
\basise{1}\wedge\basise{2} \}$.
The event's coordinates are
$\Sigma=\Sigma(ct,x,y,{\cal R}\theta)$, where the position is given
by $(x,y)$ and the scalar time needs the universal constant of the
speed of light ``$c$'' applied to convert the scale to distance units.
The bivector coordinate $\theta$ tells the angular position of the
hockey puck.  In order to have units of distance,
we need another fundamental physical constant ${\cal R}$ which we
loosely interpret as the {\it radius of gyration} (for a fundamental
particle it will be within a geometric factor of the Compton wavelength).

The {\it Clifford algebra} associated with the $(++)$ metric signature
is: ${\bf R}(2)=M(2,{\bf R})={\tt End}\ {\bf R}^{2,0}$,
isomorphic with two-by-two real matrices.  The unit bivector
$\baseE{3}=\basise{1}\wedge\basise{2}$ must then square to negative
unity.  The differential element
$d\Sigma$ and its main involution $\BAR{d\Sigma}$ are,
$$d\Sigma=dq^A\ \baseE{A} = c dt\ {\bf 1} + dx\ \basise{1}
+dy\ \basise{2} + {\cal R} d\theta\ \basise{1}\wedge\basise{2}, \eqno(5a)$$
$$\BAR{d\Sigma}= dq^A \BAR{\bf E}_A
=c dt\ {\bf 1} - dx\ \basise{1}
-dy\ \basise{2} - {\cal R} d\theta\ \basise{1}\wedge\basise{2}, \eqno(5b)$$
from which we can construct a scalar quadratic form analogous to \eq{1},
$$d\lambda^2 = \BAR{d\Sigma}d\Sigma = c^2 dt^2-dx^2 -dy^2+
{\cal R}^2 d\theta^2 
=c^2d\tau^2+{\cal R}^2 d\theta^2,  \eqno(5c)$$
which is invariant under the six parameter correlated automophism
group $O(2,2;{\bf R})$.
  
In special relativity, the affine ``proper time'' is not the same as the
ordinary time of non-relativistic space.  In polydimensional relativity,
the new affine parameter $d\lambda$ of \eq{5c} for the spinning particle
is not the same as the proper time of special relativity.  
The latter corresponds
instead to an equivalent colinear non-spinning particle.
In analogy to the introduction of the Lorentz factor \eq{3c}
to make equations relativistic, a new {\it spin correction factor}
$\Gamma$ is introduced,
$$\Gamma={d\tau \over d\lambda} = 
{\displaystyle \left({1- {{\cal R}^2  \dot{\theta}^2 
\over c^2}}\right)}^{-{1 \over 2}}=
{\displaystyle \left({1+ {{\cal R}^2\omega^2 
\over c^2}}\right)^{-{1 \over 2}}}, \eqno(6)$$
where the ``dot'' refers to differentiation with respect to the new 
affine parameter $\lambda$,
and $\omega={d\theta/d\tau}=\dot{\theta}/ \Gamma$
is the angular velocity relative to the ``old'' proper time.
In special relativity the speed of light cannot be
exceeded, here the angular velocity $\dot{\theta}$ (with respect to
parameter $\lambda$) cannot exceed $c/{\cal R}$, although $\omega$ can go
to infinity.

\subsection*{C.  Polydimensional Mechanics}
Lets continue with our 2D example of a hockey puck.  We propose a
generalization of \eq{2}, where the Lagrangian is based upon the
polydimensional form of \eq{5c}, which is invariant under the
Automorphism group $O(2,2)$,
$${\cal A}=
\int{m_0 c \sqrt{d\Sigma \BAR{d\Sigma}}} 
=\int{m_0c\ d\lambda}
=\int{m_0c\ {d\lambda \over d\tau}d\tau}=
\int{d\tau \  {m_0 c \over \Gamma} }. \eqno(7)$$
When re-parameterized in terms of the more familiar proper time using
\eq{6}, and 
compared with \eq{2} it appears as if the spin has increased the rest
mass by a factor of the inverse spin correction factor \eq{6}.  Indeed
the four-momentum derived from the Lagrangian gives the momentum:
$\vector{P}=m\vector{v}$, and energy: $E=mc^2$, where the
{\it spin-corrected \underline{linear} mass} is,
$$m={\gamma m_0 \over \Gamma} =\gamma m_0
\sqrt{1+{{\cal R}^2 \omega^2 \over c^2}}.\eqno(8)$$ 
These are physically reasonable results, however they do not agree with
the standard formulas \eq{4abcd}.  In particular, \eq{8} differs
significantly from what one might derive from standard special relativity
for the total energy of a macroscopic rotating object with center-of-mass
speed $\vector{v}$,
$$m^\prime= {\gamma m_0 \over \displaystyle 
\sqrt{1- {{\cal R}^2\omega^2 \over c^2}} }. \eqno(9)$$
Note however that this last statement is \underline{also}
not derivable from \eq{4a}.

One desirable feature of ``standard'' \eq{9} over \eq{8} is that there is
a limit on the angular velocity: $\omega={\cal R}/c$, such that the tangent
speed of the rim of the object will not exceed the speed of light.
However, the spin angular momentum (of say a ring of mass):
${\bf L}=m^\prime {\cal R}^2\omega$,
will go to infinity as the angular velocity approaches this limit.  In our
interpretation however, the angular velocity may well go to infinity,
but the angular momentum,
$${\bf L}={\delta {\cal L} \over \delta \omega}=
\Gamma m_0 {\cal R}^2 \omega, \eqno(10)$$
approaches a finite limit:
$ \displaystyle\lim_{\omega\rightarrow\infty}{\bf L}
= m_0{\cal R} c$.
The appearence is that the``rim'' speed for the bare mass approaches $c$ as
a limit as desired.  This is a very pleasing result for if we 
quantize the spin angular
momentum to be $h/4\pi$ (where h= {\it Planck's constant}), the radius of
gyration ${\cal R}$ will be the Compton wavelength (over $4\pi$).
Another interesting feature is that the spin correction to the mass
is $\left(\Gamma m_0 \right) $ in the rotational motion of \eq{10};
differing from $\left({m_0 / \Gamma}\right)$ in \eq{8} for the 
linear motion.

\section*{III. Extension of General Relativity}
Einstein's general theory of relativity requires the laws of physics
to be form invariant (covariant) under general coordinate transformations.
Physical quantities are represented by tensors, which necessarily preserve
their rank under coordinate transformations, e.g. a vector is a vector
to all observers.  Even in a curved space, under parallel transport a vector
cannot change into a bivector (nor change length, although it may twist).
In our generalization, this will no longer be the case.

\subsection*{A. Review of Standard Theory}
The {\it weak equivalence principle} states 
that the trajectory of a freely
falling body in a gravitational field is independent of its internal
structure and composition (e.g. heavy balls fall just as fast as
light ones).  The
{\it strong equivalence principle} states that an accelerated reference
frame is equivalent to gravitation, or that mass curves space, and 
accelerated motion is due to the curvature.

In general coordinates, the tangent basis vectors:
$\basise{\mu}=\partial_\mu \Sigma$ at event $\Sigma$
are a function of the coordinates.  Under differential
displacement the basis vectors change,
$$\partial_\alpha \basise{\mu}=\Gamma_{\alpha\mu}^{\ \ \beta}
\ \basise{\beta}, \eqno(11)$$
where in a space without torsion the affine connections are symmetric
in the lower indices: $\Gamma_{\alpha\mu}^{\ \ \beta}=
\Gamma_{\mu\alpha}^{\ \ \beta}$.
The generalization of \eq{1} requires the introduction of
the {\it metric tensor}  $g_{\alpha\beta}$, which contains all the
information necessary to describe gravitation,
$$c^2d\tau^2=dx^\alpha dx^\beta \ g_{\alpha\beta}, \eqno(12a)$$
$$ g_{\alpha\beta}=\basise{\alpha}\cdot \basise{\beta}
={1 \over 2}\{\basise{\alpha},\basise{\beta}\}. \eqno(12b)$$
The latter equation is the definition of a Clifford algebra in general
coordinates.
The differential of the metric tensor can hence be computed directly from
\eq{12b} and \eq{11} \underline{only if}
 the Leibniz rule for differentiation holds.
While not generally true (e.g. in a Weyl space), it 
is the condition for a Riemann space,
$$\partial_\mu g_{\alpha\beta}=(\partial_\mu\basise{\alpha})\cdot
\basise{\beta}+\basise{\alpha}\cdot(\partial_\mu \basise{\beta})
\\=\Gamma_{\mu\alpha}^{\ \ \delta}\ g_{\delta\beta}+
\Gamma_{\mu\beta}^{\ \ \delta}\ g_{\delta\alpha}. \eqno(13)$$
By permutation, one can solve for the affine connections in terms
of the metric tensor.

Trajectories in curved space can be derived from the action integral
of \eq{2} by substituting \eq{12a} for the proper time.  The result is
known as the {\it geodesic equation}, which describes the shortest
path between two points in curved space,
$$\ddot{x}^\mu = - \dot{x}^\alpha \dot{x}^\beta 
\Gamma_{\alpha\beta}^{\ \mu}. \eqno(14)$$
This is consistent with the weak equivalence principle, in that
all particles follow the same path independent of mass (e.g. big balls
fall at the same rate as small balls).
In a Riemann space, the parallel displacement of a vector over a small closed
loop will not change its length, but may rotate the vector in proportion
to the amount of curvature (due to gravity),
$$\Delta V^\nu = R_{\alpha\beta\mu}^{\ \ \ \nu}\ V^\mu\ \Delta
A^{\alpha\beta}, \eqno(15a) $$
$$R_{\alpha\beta\mu\nu}=\basise{\beta}\cdot [\partial_\mu,\partial_\nu]
\basise{\alpha}, \eqno(15b)$$
where $\Delta A^{\alpha\beta}$ is the oriented area of the loop, and 
$R_{\alpha\beta\mu\nu}$ is the Riemann curvature tensor.

In a Weyl space however, the \underline{length} of a vector 
\underline{can} change under parallel
displacement.  The Leibniz rule is no longer valid, such that \eq{13}
no longer holds.  It is replaced by fiat with,
$$\partial_\mu \ g_{\alpha\beta}=\Gamma_{\mu\alpha}^{\ \ \delta}
\ g_{\delta\beta}+
\Gamma_{\mu\beta}^{\ \ \delta} \  g_{\delta\alpha}
+\phi_\mu \  g_{\alpha\beta}, \eqno(16)$$
where $\phi_\mu$ was originally intended by Weyl\cite{Weyl} to be
the electromagnetic vector potential, however the approach did not yield
the correct electrodynamic equations.  The parallel transport of a
scalar (such as length of vector $V^2$) around a closed loop would yield:
$\Delta V^2 =V^2\ F_{\mu\nu}\ \Delta A^{\mu\nu}$, where:
$F_{\mu\nu}=\partial_\mu \phi_\nu - \partial_\nu \phi_\mu$.

It has been argued by Papapetrou\cite{Papa} that a fully covariant
equation of motion for a spinning particle would differ from \eq{14},
$$\dot{p}^\mu =-p^\alpha \dot{x}^\beta 
\Gamma_{\alpha\beta}^{\ \ \mu} -{1 \over 2} 
R_{\rho\sigma}^{\ \ \mu\nu}\ \dot{x}_\nu \ S^{\rho\sigma}, \eqno(17)$$
where the spin tensor is still given by \eq{4c} and the
momentum by \eq{4a}.  While a non-spinning particle will follow a
geodesic, a spinning one will travel a different path, which
clearly violates the weak equivalence principle.

\subsection*{B.  Polydimensionally Affine Space}
The tangent basis multivectors ${\bf E}_A=\partial_A \Sigma$ of the
polydimensional Clifford manifold are functions of the full set of
generalized coordinates,
$${\bf E}_A(q^B)=\Delta_{A}^{\ B}(q^C)\ \widehat{\bf E}_B. \eqno(18)$$
We call $\Delta_{A}^{\ B}$ the {\it geobeins}\cite{PezzMex}
(``geometry legs''), which are
completely analogous to Crawford's {\it drehbeins}\cite{Crawford}
except here we are reshuffling \underline{physical}
geometry at every point.  The 
fiducial basis $\widehat{\bf E}_A$ is assumed to be the Clifford group
generated by an orthonormal basis which satisfies \eq{12b}.  
However, \eq{12b} will no hold for the generalized tangent basis
vectors: $\basise{\alpha}$ unless the geobeins are severely restricted.
For example, \eq{12b} can not accomodate an
idempotent/nilpotent basis which does not have an identity element.

The general form which would allow for that possibility could be
expressed as a {\it Jordan algebra}: $\{ {\bf E}_A,{\bf E}_B \}
=2 {\cal G}_{AB}^{\ \ C} {\bf E}_C $, where ${\cal G}_{AB}=
{\cal G}_{AB}^{\ \ 0}{\cal G}_{00}^{\ \ 0}$ would be the 
{\it Cartan metric}.  However, for the purposes of this paper
we propose the mild generalization as an ansatz,
$${1 \over 2} \{\basise{\alpha},\basise{\beta}\}=g_{\alpha\beta}
\ \Ed{0}, \eqno(19a)$$
which among other enhancements, generalizes \eq{12b} to include
Weyl space.  We further propose the simplifying restriction
that the basis scalar $\Ed{0}$ commutes locally with all elements, and
$\{\Ed{0},\Ed{0}\}=2 g_{00} \Ed{0}$,
where $g_{00}$ is the ``scale'' or metric of the scalar coordinate.
We assume that the wedge product of basis vectors is still
given by the Lie product:
$[\basise{\alpha},\basise{\beta}]=
2\basise{\alpha}\wedge\basise{\beta}$,
so that the Clifford product of two basis vectors may now be written,
$$\basise{\alpha}\basise{\beta}=g_{\alpha\beta}\ \Ed{0}
+\basise{\alpha}\wedge\basise{\beta}. \eqno(19b)$$

The generalized {\it polydimensional connection}
$\Lambda_{AB}^{\ \ C}$ is defined,
$${\partial{\bf E}_A \over \partial q^B}=\Lambda_{AB}^{\ \ C}
\ {\bf E}_C. \eqno(20)$$
In a Clifford manifold, the basis elements are interdependent; a bivector
is the outer product of two basis vectors.  Hence the connection of a
multivector may be derived from the connections of the basis vectors.
Note however that the {\it Leibniz rule is no longer valid for the
inner (dot) or outer (wedge) products} because the definitions of these
products involved an alternating sign depending upon the rank of the
geometry, which is no longer fixed.  {\it The Leibniz rule is
however valid for the Clifford direct geometric product}.

At this point we take an epagoic approach by using simple examples to
illustrate the new features.  Let us return to our 2D ``hockey puck''
problem.
The explicit form of \eq{20} for the two basis vectors is,
$$\partial_A \basise{\mu}=\sigma_{A\mu}\Ed{0}+
\Gamma_{A\mu}^{\ \ \nu}\ \basise{\nu} +\lambda_{A\mu}\ \Edd. \eqno(21a)$$
Then the connection for the bivector $\Ed3 = \Edd$ is hence
completely determined from \eq{21a},
$$\partial_A \left(\Edd \right)=\HALF [\partial_A \Edb, \Edc] 
+\HALF  [\Edb, \partial_A\Edc]
=\Gamma_{A\alpha}^{\ \ \alpha}\ \Ed3 + Q_A^{\ \nu}\ \basise{\nu}, \eqno(21b)$$
$$Q_A^{\ 1}=g_{00}\ \left( \lambda_{A1}\ g_{22} 
-\lambda_{A2}\ g_{12} \right), \eqno(21c)$$
$$Q_A^{\ 2}=g_{00}\ \left( \lambda_{A2}\ g_{11}
-\lambda_{A1}\ g_{12} \right). \eqno(21d)$$
The connection for the basis scalar,
$$\partial_A \Ed{0}=-\phi_A \Ed{0}+M_A^{\ \mu}\ \basise{\mu}
+N_A\ \Edd, \eqno(21e)$$
can be simplified by differentiating \eq{19a}.  Equating terms of
similar geometry, the bivector terms give us that $N_A =0$ under our
restrictions.  Further the scalar portion recovers \eq{16} showing
that $\phi_A$ is Weyl's connection coefficient.  The vector part shows,
$$M_A^{\ \mu}\ g_{\mu\delta}=\sigma_{A\delta}\ g_{00}. \eqno(21f)$$

The parallel displacement of a vector around a closed loop could now return
as a completely different object (e.g. a bivector).  One would
generalize the curvature formula \eq{15b} to something like,
$$[\partial_A,\partial_B]\Ed{C}=
{\cal F}_{ABC}^{\ \ \ D}\ \Ed{D}. \eqno(22a)$$
In our 2D case, a loop in the x-y plane would yield something like,
$$[\partial_\mu,\partial_\nu]\ \basise{\alpha}=
R_{\mu\nu\alpha}^{\ \ \ \beta}\ \basise{\beta}+
W_{\mu\nu\alpha}\Ed{0}+V_{\mu\nu\alpha}\ \Edd, \eqno(22b)$$
Now this becomes more acceptable if you start out with objects that
are multivectorial in the first place; in fact we propose that particles
have scalar$+$vector$+$bivector parts to represent their mass,
linear motion, spin, etc.  The curvature which bends one type of
geometry into another is simply a coupling of these various portions
(e.g. a spin contribution to linear momentum).
Even more strange however is that we can have closed paths which are
not in the ordinary vector coordinates, but involve the coordinates 
associated with the other basis multivectors.  Hence a particle which
is ``spun'' then translated will be in a different state than one which
is translated then spun.  We can even have 
multivector paths, which are not just one-dimensional
lines, but part scalar, part linear and part area.

\subsection*{C.  Polygeometrodynamics}
We generalize by proposing a new equivalence principle, that the laws of
physics should be fully covariant under \underline{local}
automorphism transformations.  Generalized forces will be associated
with curvature which bends one type of geometry into another
(e.g. vector twisted into scalar).

It remains to be shown that the connection coefficients can be derived
from some sort of generalized metric (e.g. the Cartan metric).  Further
it would be nice to have some generalized form of the action integral
\eq{7} from which the equations of motion can be derived.  By induction
we believe that with the proper development one obtains 
generalized {\it polygeodesics}, which resemble \eq{14}:
$\ddot{q}^A+\dot{q}^B \dot{q}^C \Lambda_{BC}^{\ \ A}=0$,
where the differentiation is in terms of the affine parameter 
$\lambda$ which is defined by the generalization of \eq{5c}
to polydimensionally curved space.

We present a simplified case, where there is no change in the scale,
such that \eq{21e} is zero, hence $\sigma_{A\mu}$ in \eq{21a}
also vanishes. The geodesics are of the form,
$$\ddot{x}^\nu=-\dot{x}^\alpha \dot{x}^\beta 
\Gamma_{\alpha\beta}^{\ \ \nu}-\left({\cal R}\dot{\theta}\right)^2
Q_3^{\ \nu}-{\cal R}\dot{\theta}\dot{x}^\alpha 
\left( Q_\alpha^{\ \nu} + \Gamma_{3\alpha}^{\ \nu} \right),\eqno(23a)$$
$${\cal R}\ddot{\theta}=-{\cal R}\dot{\theta} \dot{x}^\beta 
\left( \Gamma_{\beta\alpha}^{\ \ \alpha}+\lambda_{3\beta}\right)
-\dot{x}^\alpha \dot{x}^\beta \lambda_{\alpha\beta}
-\left( {\cal R}\dot{\theta}\right)^2 
\Gamma_{3\alpha}^{\ \alpha}, \eqno(23b)$$
where the subscript $3$ is associated with the spin coordinate:
$q^3={\cal R}\theta$.  The second equation shows that the spin 
geodesic has a new torque proportional to the linear motion coupled by 
$\lambda_{\alpha\beta}$.  Comparing the first equation to the
Papapetrou equation (17) suggests that we might want to make
the identification: $2\Gamma_{3\mu}^{\ \nu}={\cal R} R_{12\mu}^{\ \ \nu}$,
which implies,
$${\partial \over \partial \theta}={{\cal R}^2 \over 2}
[\partial_1 , \partial_2 ]. \eqno(24)$$
In other words, the commutator derivative of general relativity
might be equivalent to differentiation with respect to a bivector.

\section*{IV.  Summary of Principles}
We summarize our explorations epagogically, by proposing several broad
organizing principles.  Just as tensors were the natural language in 
which to formulate general relativity, Clifford algebra is the natural
language in which to express the polydimensional theory.

{\bf Principle of Relative Dimension.}  In standard relativity, a
scalar (point) is the same to all observers, in all coordinate systems.
While a line may be bent due to curvature, its length is unchanged.
Now {\it Dimension is in the eye of the beholder}.  The geometric rank
that an observer assigns to an object (e.g. bivector) is a function of
the observer's frame.  It might be possible to logically extend this 
statement to say that there is no absolute dimension to the universe.

{\bf Polydimensional Isotropy.}  `No preferred direction' is extended to
mean that there is no absolute direction to which you can assign the geometry
of a vector.  For example, if we turn out the lights and exchange the 
basis vectors for their dual trivectors in all formulas in 4D, you can't
tell that a change was made.

{\bf The Greider Maxima.}  To be complete, the laws of physics must
be multivectorial in form (having scalar, vector, bivector etc. parts).
\underline{Every} geometric piece of a 
multivector equation must be physically
interpretable.  A separate `Spin space' is an unneeded construct.

{\bf Polydimensional Covariance.}  The laws of physics should be
form invariant under local automorphism transformations, which reshuffle
the physical geometry.  Spin gauge theory (in spinor space) is not therefore
an artifact of spin space, it is a manisfestation of this broader
classical principle.

\section*{V. Acknowledgements}
In particular we thank W. Baylis, University of Windsor, Ontario,
Canada, for the invitation in the summer of 1994, from which these
ideas unfolded.  C. Doran's sarcastic comments (after numerous libations)
at the Banff Summer School 1995
{\it (will the parallel transport of a vector around a closed loop turn
it into a bottle of port?)} goaded me into coming up with a simple
example to prove my case.
Finally we thank J. Crawford (Penn. State U.) for his
careful reading of the early draft.

\end{document}